# Steerable sound transport in a 3D acoustic network

Bai-Zhan Xia*, Jun-Rui Jiao, Hong-Qing Dai, Sheng-Wen Yin, Sheng-Jie Zheng, Ting-Ting Liu, Ning Chen, De-Jie Yu*

State Key Laboratory of Advanced Design and Manufacturing for Vehicle Body, Hunan University, Changsha, Hunan, People's Republic of China, 410082

Quasi-lossless and asymmetric sound transports, which are exceedingly desirable in various modern physical systems, are almost based on nonlinear or angular-momentum biasing effects with extremely high power levels and complex modulation schemes. A practical route for the steerable sound transport along any arbitrary acoustic pathway, especially in a 3D acoustic network, could revolutionize the sound power flow and the sound communication. Here, we design an acoustic device consisting of a regular-tetrahedral cavity with four cylindrical waveguides. A smaller regular-tetrahedral solid in this cavity is eccentrically emplaced to break its spatial symmetry. The numerical and experimental results show that the sound power flow can unimpededly transport between two waveguides away from the eccentric solid within a wide frequency range. Furthermore, in the vicinity of eigenmode, the sound waves from various waveguides can transport to different waveguides, respectively along the compressed and broadened pathways without mutually interference. Endowed with these quasi-lossless and asymmetric transport characteristics, we construct a 3D acoustic network, in which the sound power flow can flexibly propagate along arbitrary sound pathways defined by our acoustic devices with eccentrically-emplaced regular-tetrahedral solids.

xiabz2013@hnu.edu.cn (Baizhan Xia)

djyu@hnu.edu.cn (DejieYu)

Phononic crystals or acoustic metamaterials, made of periodic composite materials and structures, have exhibited outstanding abilities to manipulate the sound propagation, from the negative refraction[1-7], the high blocking[8-10], the acoustic supertunneling[11-15] to the topological one-way transport[16-28]. However, a highly promising property, namely the reconfigurable sound transport in a 3D acoustic network, has been evaded in periodic acoustic systems. The symmetry breaking which is uncommon in nature is deliberately created in various artificial systems for exotic physical properties. In the plasmonic metamaterial, the higher order plasmon resonance activated by the symmetry breaking can lead to novel material functionalities, such as long lifetimes and distinct absorption losses[29]. Phononic crystals and acoustic metamaterials with the inversion symmetry breaking produced unidirectional transmissions[30-36]. By breaking the coupling symmetry between discrete closed metallic nano-rings, the optical metamaterial yielded an ultrabroad band with negative index[37]. The macroscopic symmetry breaking in paraelectric phases resolved the argument about the large flexoelectric coefficients of ferroelectric perovskite materials[38, 39] and was of interest for nanoelectronics[40, 41]. The chiral magnetoelastic metamaterial with the spontaneous symmetry breaking leaded to an artificial phase transition[42]. Subsequently, the spontaneous symmetry breaking in the coupled photonic-crystal nanolaser[43], in the split potential box[44] and in the quasi-double-well potential[45] have been deeply investigated. The multiple scattering induced by the spatial symmetry breaking produced a negative refraction, an acoustic superlensing and a high reflection[6, 10, 46]. The symmetry-breaking transition of a curved multilayered surface has been successfully employed to determine elastic surface patterns[47]. Massless Dirac fermions and quasiparticles have emerged in graphene lattices with symmetry breaking[48-50].

Recently, Dai et al.[51] realized a quasi-lossless sound transport along any arbitrary pathway in a 2D acoustic network consisted of phononic metacrystals with the spatial symmetry breaking. The robust sound propagation along edges of 2D phononic lattices with the time-reversal symmetry breaking[16-21] have been systematically investigated. Up to now, it is still a great challenge to configure a sound pathway in a 3D acoustic network, even though the topological edge states were creatively realized in phononic systems[16-21]. Compared with the 2D acoustic network[16-18, 52], the 3D acoustic network will exhibit a much more flexible ability to control the sound propagation and provide a much more extraordinary platform to design the noticeable acoustic device.

In this paper, we design an asymmetric acoustic device to steer the quasi-lossless sound transport in a 3D acoustic network. The acoustic device is consisted of a regular-tetrahedral cavity with four cylindrical waveguides, shown in Fig. 1a. The side length of the regular-tetrahedral cavity is 220mm. The inside diameter and the length of the cylindrical waveguide are 40mm and 150mm. The wall thicknesses of the cavity and waveguides are 5mm. A regular-tetrahedral solid is emplaced in the cavity. The side length of this regular-tetrahedral solid is 110mm. Our acoustic device is made of the stainless steel. Surfaces and edges of the regular-tetrahedral solid are respectively parallel to those of the regular-tetrahedral cavity. The regular-tetrahedral solid can freely shift in the plane crossing the edge and center of the regular-tetrahedral cavity. The eccentricity of the regular-tetrahedral solid is defined as the distance between $O_1$ and $O_2$. Due to the eccentric emplacement of the regular-tetrahedral solid, interspaces between various surfaces of the regular-tetrahedral solid and cavity are different. When the regular-tetrahedral solid approaches to the edge $N_1$ of the regular-tetrahedral cavity

(illustrated in Fig. 1a), interspaces between surfaces $L_1$ and $S_1$ and between $L_4$ and $S_4$ are compressed. On the contrary, interspaces between surfaces $L_2$ and $S_2$ and $L_3$ and $S_3$ are broadened.

Full-wave numerical simulations are performed to investigate the excellent transport properties of our acoustic device. The sound transmission spectra of the symmetric acoustic device and the asymmetric one are illustrated in Fig. 1b and c, assuming that the sound waves are incident from the waveguide $W_1$. The eccentricities for the symmetric acoustic device and asymmetric one are respectively 0mm and 25mm. In the first case, the sound power flow is evenly divided to three waveguides $W_2$, $W_3$ and $W_4$, which are symmetrically placed around the axis line of the waveguide $W_1$. As most of the sound power is reflected back by the centrically-emplaced regular-tetrahedral solid, the sound power divided to each of three waveguides is weak. In the second case, the sound power can unimpededly propagate between the waveguides $W_1$ and $W_2$ with a high transmission, even up to 100%. Namely, almost all sound power flow is steered to the waveguide $W_2$ in the frequency range between two eigenmodes 3836.4Hz and 4252.4Hz, leaving the extremely weak sound power at the waveguides $W_3$ and $W_4$. The reason is that when the regular-tetrahedral solid is deviated from the center of the acoustic device, the symmetries of eigenmodes (shown in Fig. 1e and 1f) are broken. Based on the constructive coupling between the waveguides $W_1$ and $W_2$ arising from the scattering effect of the eccentrically-emplaced regular-tetrahedral solid, a strong pressure field distribution between them is created. On the contrary, due to the destructive uncoupling of the other pairs of waveguides, a null of pressure field at the waveguides $W_3$ and $W_4$ is presented. Compared with the other approaches for the quasi-lossless sound transport with a narrow frequency range, the unimpeded sound propagation in our acoustic device can be efficiently done in a relative bandwidth of over 10% without any additional real-time modulation. When the regular-tetrahedral solid is removed from the acoustic device, the sound power flow cannot transport among these waveguides, even if two waveguides are blocked (in Fig. S1 of the Supplementary Information). As a result, by introducing an eccentrically-emplaced regular-tetrahedral solid, our asymmetric acoustic device exhibits an excellent transport between the waveguides located at the opposite direction of offset, which is not implemented in conventional acoustic devices.

Furthermore, if the sound wave is incident from the waveguide $W_3$, the sound power flow, in vicinities of two eigenmodes, can efficiently transport to the waveguide $W_4$, but not the waveguides $W_1$ and $W_2$, as shown in Fig. 1d. Synthesizing Fig. 1c and 1d, we can observe an excellent phenomenon that the sound wave from the waveguide $W_1$ can only transport to the waveguide $W_2$ through the broadened pathway, while the sound wave from the waveguide $W_3$ can only transport to the waveguide $W_4$ through the compressed pathway (in Fig. S2 of the Supplementary Information). Thus, the sound waves from the waveguides $W_1$ and $W_3$ can respectively transport to the waveguides $W_2$ and $W_4$ without interfering with each other, providing an extraordinary platform for the sound power transmission and the sound signal communication.

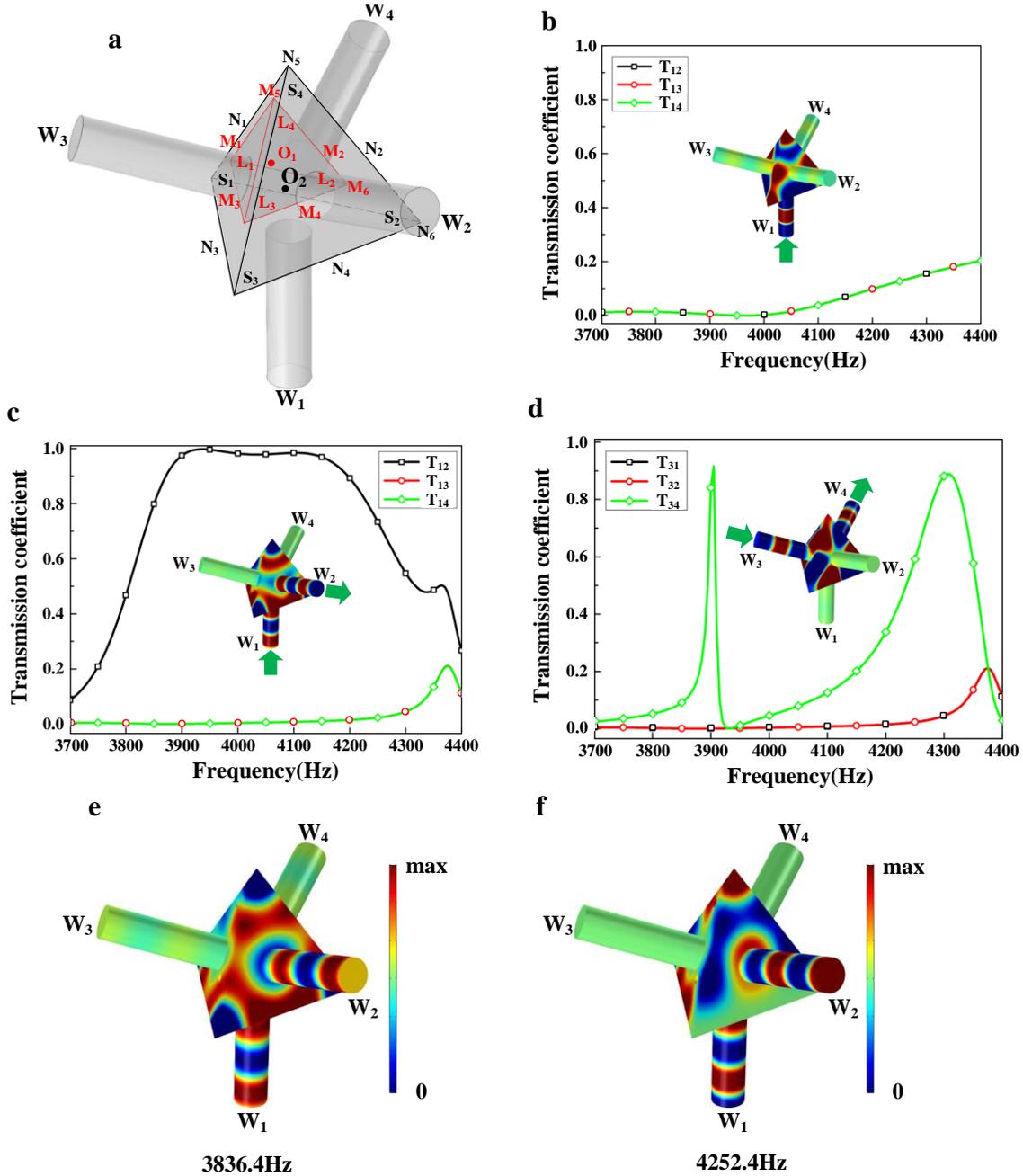

**Figure 1 | Schematic diagram, sound transmission spectra and eigenmodes of our acoustic device. a**, The geometry of our acoustic device. $L_1$, $L_2$, $L_3$ and $L_4$ are surfaces of the regular-tetrahedral solid. $S_1$, $S_2$, $S_3$ and $S_4$ are surfaces of the regular-tetrahedral cavity. $N_1$, $N_2$, $N_3$, $N_4$, $N_5$ and $N_6$ are edges of the regular-tetrahedral cavity. $M_1$, $M_2$, $M_3$, $M_4$, $M_5$ and $M_6$ are edges of the regular-tetrahedral solid. $O_1$ and $O_2$ stand centers of solid and cavity. **b** and **c,** The sound transmission spectra of the symmetric acoustic device and the asymmetric one when the sound waves are incident from the waveguide $W_1$. **d,** The sound transmission spectra of the

asymmetric acoustic device when the sound waves are incident from the waveguide $W_3$. Acoustic pressure field distributions at 3900Hz are inserted in these sound transmission spectra. $T_{12}$, $T_{13}$ and $T_{14}$ represent the transmission spectra of sound from the waveguide $W_1$ to the waveguides $W_2$, $W_3$ and $W_4$. $T_{31}$, $T_{32}$ and $T_{34}$ represent the transmission spectra of sound from the waveguide $W_3$ to the waveguides $W_1$, $W_2$ and $W_4$. **e** and **f,** Eigenmodes of the asymmetric acoustic device at resonant frequencies 3836.4Hz and 4252.4Hz.

To gain convincing insights into behaviors of our acoustic device, the magnitude transmissions at these waveguides are measured in experiments. Photograph of the fabricated acoustic device is presented in Fig. 2a. When the sound wave is incident from the waveguide $W_1$, the magnitude transmissions of sound at the waveguides $W_2$, $W_3$ and $W_4$ in our acoustic device with an eccentrically-emplaced regular-tetrahedral solid are illustrated in Fig. 2b. As predicted, the magnitude transmission of sound to the waveguide $W_2$ is much larger than that to the waveguides $W_3$ and $W_4$ in the considered frequency range between two eigenmodes. This indicates that almost all sound energy is steered to the waveguide $W_2$, which is excellently agree with our numerical results presented in Fig. 1c. The effect of the varying eccentricity on the magnitude transmission of sound from the waveguide $W_1$ to the waveguide $W_2$ is presented in Fig. 2c. When the eccentricity is 5mm, the amplitude transmission of sound to the waveguide $W_2$ is less, especially in the vicinity of the lower eigenmode 3836.4Hz. With the increase of the eccentricity, the amplitude transmission of sound to the waveguide $W_2$ gradually increases until the eccentricity is up to 20mm. When the eccentricity is larger than 20mm, the amplitude transmission of sound almost converges, even if the eccentricity further increases. This makes our acoustic device immunize against the fluctuation of eccentricity when it is larger than

20mm. In other words, our acoustic device exhibits an extraordinarily robust property over a moderately broad range of eccentricity, from 20mm to 31.82mm. This strongly robust capability is an additional advantage of our acoustic device compared with conventional acoustic devices which are sensitive to the uncontrollable variations, such as manufacturing errors.

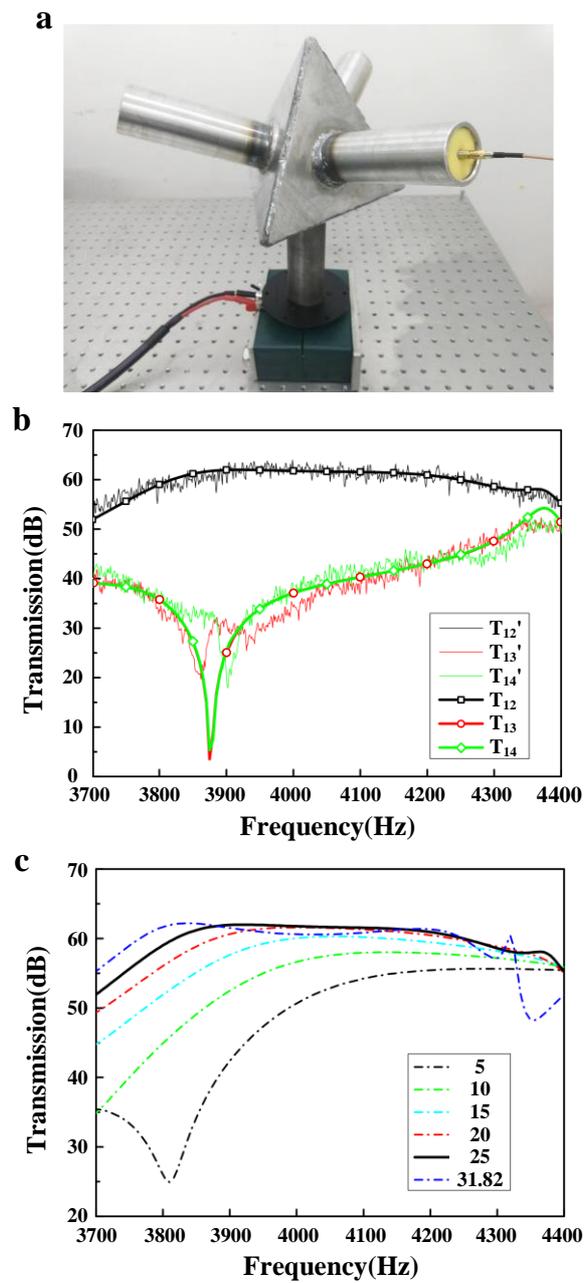

**Figure 2 | Transmission coefficients. a,** Photograph of the fabricated acoustic device. **b,** Magnitude transmissions of the asymmetric acoustic metacrystal. $T_{12}'$, $T_{13}'$ and $T_{14}'$ represent the measured magnitude transmissions from the waveguide $W_1$ to the waveguides $W_2$, $W_3$ and $W_4$. $T_{12}$, $T_{13}$ and $T_{14}$ represent the simulated magnitude transmissions from the waveguide $W_1$ to the waveguides $W_2$, $W_3$ and $W_4$. **c,** The effect of the varying eccentricity on the magnitude transmission of sound from the waveguide $W_1$ to the waveguide $W_2$.

By deliberately offsetting the regular-tetrahedral solids, we can construct a 3D acoustic network with any arbitrary harped 3D contour defined by the asymmetric acoustic devices. Thus, compared with the fixed phononic network with a passive sound pathway, our acoustic device not only expresses a new way to realize a 3D asymmetrical transport, but also provides a well-controlled platform to reconfigure 3D sound pathways. In this report, we construct a 3D acoustic network consisting of 20×20×20 acoustic devices to verify the extraordinary transport property. Fig. 3a shows that the sound power flow at a frequency of 3900Hz can transport along edges of the 3D acoustic network with a transmission coefficient about 93.48%. Fig. 3b shows that the sound power flow can pass through the ladder-like diagonal line with a transmission coefficient about 99.08%. Fig. 3c shows that the sound power flow can pass through a complicated pathway, along which the regular-tetrahedral solids in the acoustic devices are randomly offset, with a transmission coefficient about 97.32%. The transmission spectra in those three cases, presented in Fig. S3 in the Supplementary Information, also verify that the excellent transport property of our 3D acoustic network can be realized in a relative bandwidth. Thus, with the support of our acoustic device, the sound power flow can be unimpededly steered along any arbitrary pathway defined by the asymmetric acoustic devices, confirming the possibility of dynamically-controllable quasi-lossless routing of sound in a 3D acoustic network.

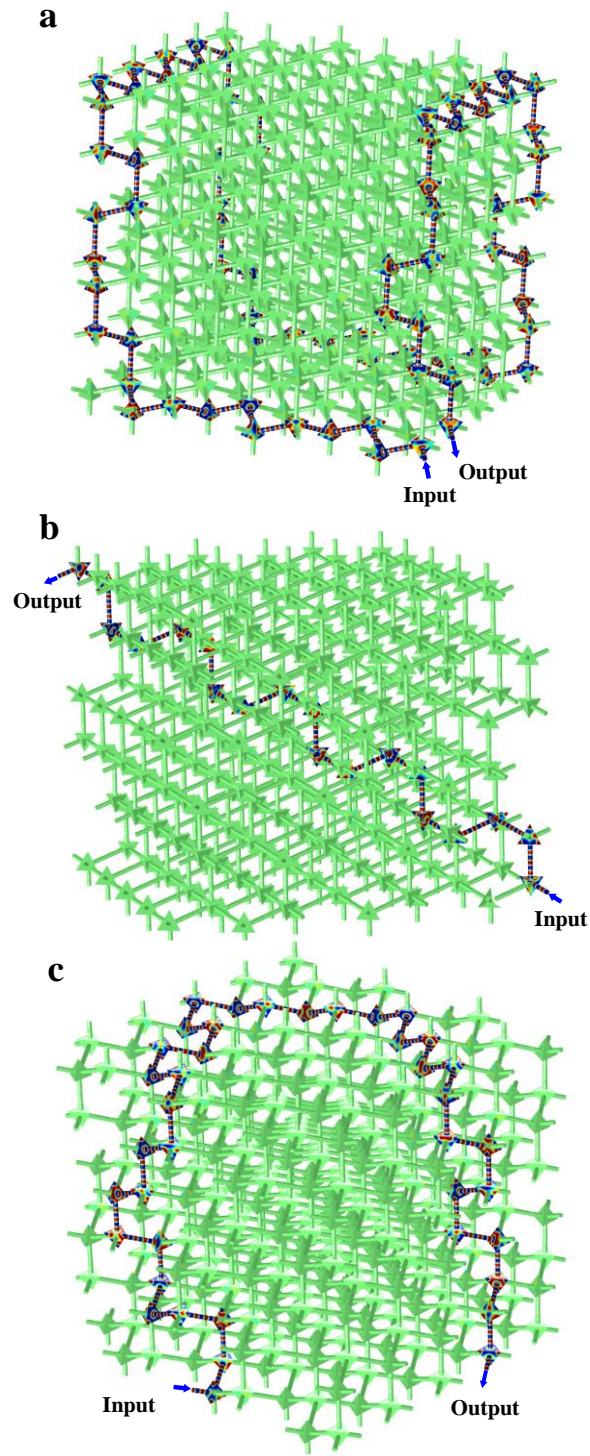

**Figure 3 |Pressure field distributions of the sound power flow within a 3D acoustic network at a frequency of 3900Hz. a,** The sound power flow transporting along edges of the 3D acoustic network. **b,** The sound power

flow passing through the ladder-like diagonal line. **c,** The sound power flow propagating along a random pathway.

In conclusion, by eccentrically emplacing the regular-tetrahedral solid, an asymmetric acoustic device is designed such that the sound power flow can unimpededly transport through the waveguides away from the eccentric solid. This extraordinary transport property is robust over a moderately broad range of eccentricity. Furthermore, inspired by this unique transmission phenomenon, we construct a 3D acoustic network in which the sound power flow can be effectively steered along arbitrary pathways such as the edge, the ladder-like diagonal line and even the random sound pathway. Thus, our acoustic devices provide an excellent platform to design 3D acoustic network with reconfigurable and quasi-lossless sound pathways. Furthermore, our findings are not limited to the audible sound. By modulating the size of model, our acoustic device can be applied to different frequencies, from audible sound to ultrasound, and even up to hypersound. By integrating with other wave systems, our finding may exhibit an extraordinary potential in steering the quasi-lossless transport of the electromagnetic wave, the light and the heat.

**Acknowledgments**

The paper is supported by National Natural Science Foundation of China (No.11402083, 11572121), Collaborative Innovation Center of Intelligent New Energy Vehicle and the Hunan Collaborative Innovation Center of Green Automobile.

# Supplementary Information

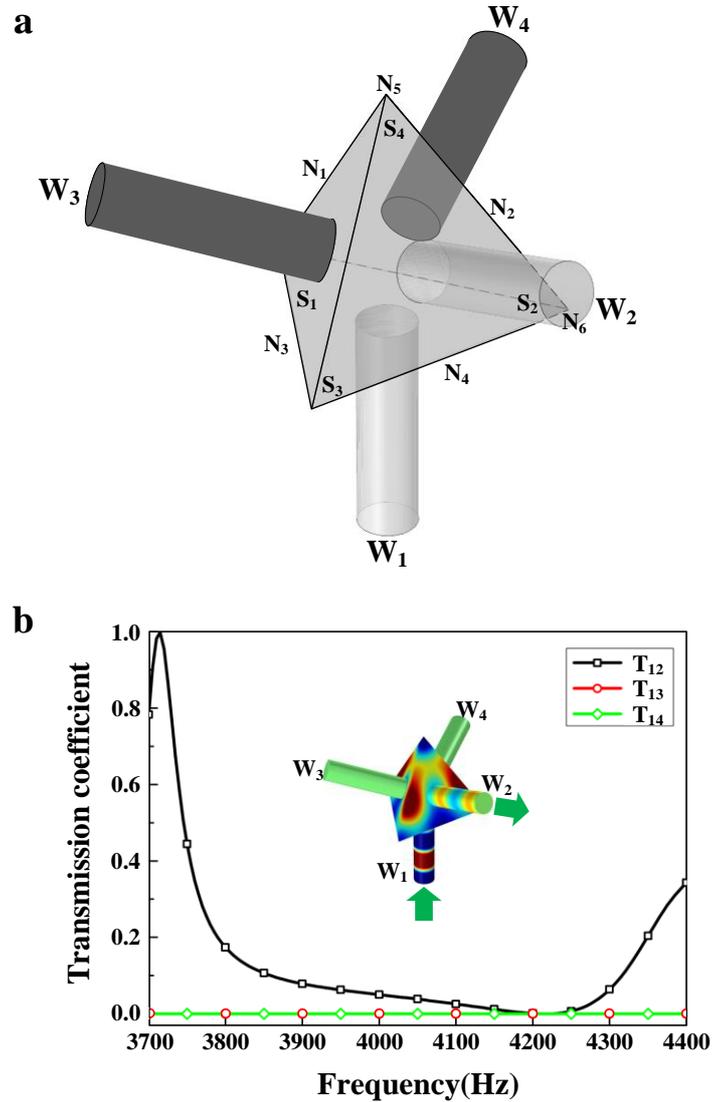

Fig.S1. **a**, The geometry of the acoustic device without the regular tetrahedral solid. The waveguides $W_3$ and $W_4$ are blocked. **b**, Transmission coefficients of the acoustic device without the regular tetrahedral solid. The sound wave is incident from the $W_1$. $T_{12}$, $T_{13}$ and $T_{14}$ represent the transmission spectra from the waveguide $W_1$ to the waveguides $W_2$, $W_3$ and $W_4$. The sound pressure field distribution at 3900Hz are inserted.

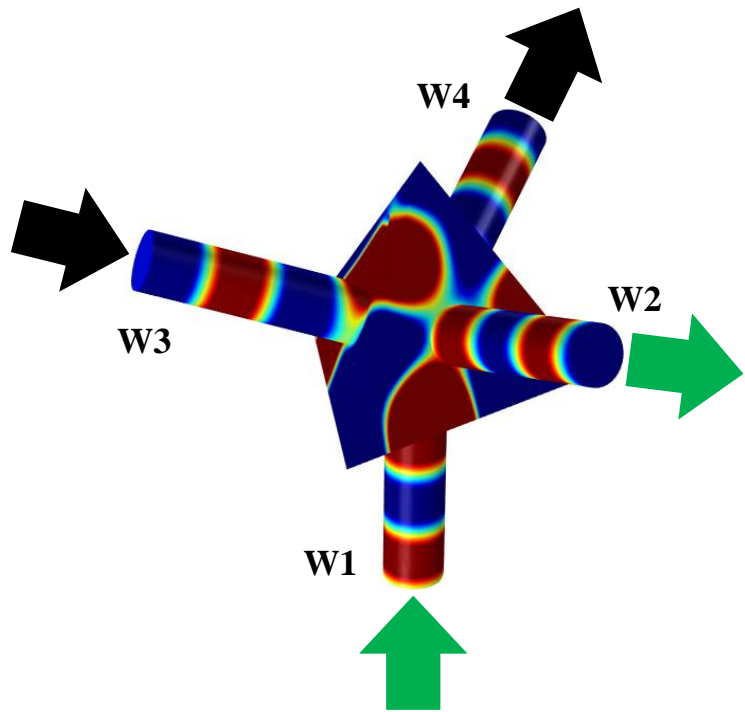

Fig.S2. The sound pressure field distribution of our asymmetric acoustic device at 3900Hz. Sound waves from the waveguides $W_1$ and $W_3$ respectively transport to the waveguides $W_2$ and $W_4$.

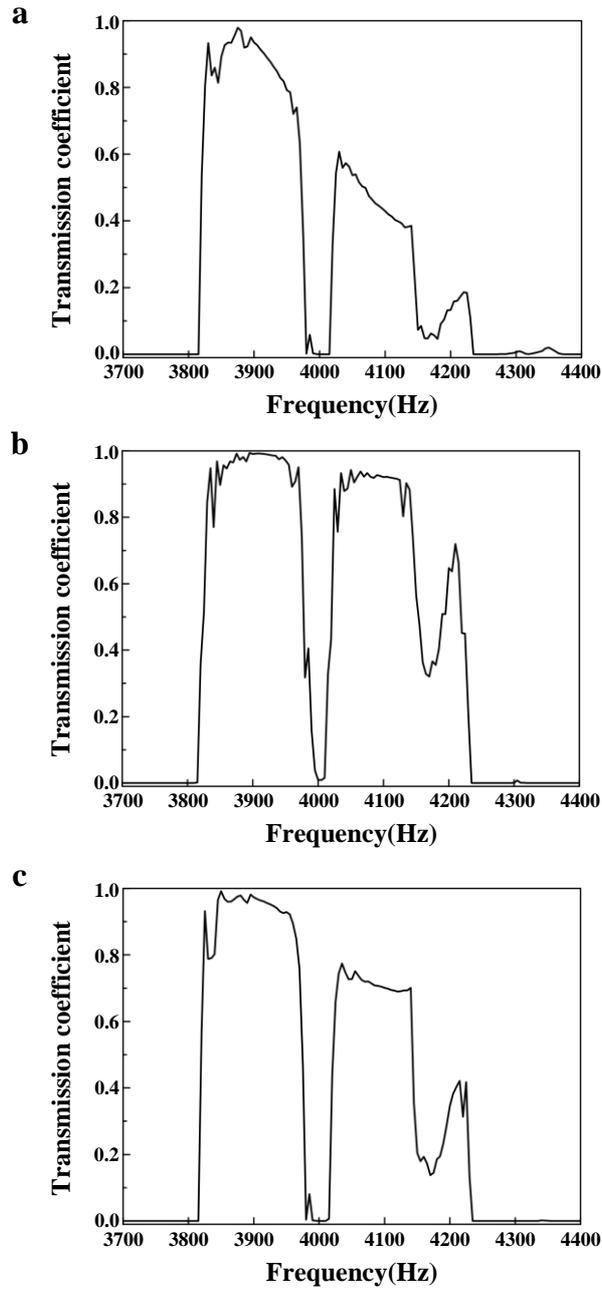

Fig.S3. Transmission coefficients of the sound power flow within a 3D acoustic network. **a,** The sound power flow transporting along edges of the 3D phononic network. **b,** The sound power flow passing through the ladder-like diagonal line. **c,** The sound power flow propagating along a random pathway.